\documentclass[draft,eqsecnum,nofootinbib,aps]{revtex4}
\begin{document}
\title{Model independent sum rules for strange form factors}
\author{Soon-Tae Hong}
\email{soonhong@ewha.ac.kr}
\affiliation{Department of Science
Education, Ewha Womans University, Seoul 120-750 Korea}
\date{September 7, 2004}

\vskip3.0cm
\begin{abstract}
We study chiral models with SU(3) group structure such as Skyrmion 
and chiral bag to yield theoretical predictions of proton strange form factor 
comparable to the recent experimental data of the SAMPLE Collaboration.  For these 
predictions we formulate model independent sum rules for proton strange form factor 
in terms of baryon octet magnetic moments.  We also investigate the 
Becci-Rouet-Stora-Tyutin symmetries associated with the St\"uckelberg coordinates, 
ghosts and anti-ghosts involved in the Skyrmion model.
\end{abstract}
\maketitle

\section{Introduction}

There have been many interesting developments concerning the strange flavor 
structures in the nucleon and the hyperons.  Especially, the internal 
structure of the nucleon is still a subject of great interest to both
experimentalists and theorists.  In 1933, Frisch and Stern~\cite{stern33} 
performed the first measurement of the magnetic moment of 
the proton and obtained the earliest experimental evidence for the internal 
structure of the nucleon.  However, it wasn't until 40 years later that the 
quark structure of the nucleon was directly observed in deep inelastic 
electron scattering experiments and we still lack a quantitative theoretical 
understanding of these properties including the magnetic moments.

Recently, the SAMPLE Collaboration reported the
experimental data of the proton strange magnetic form factor through parity violating
electron scattering at a small momentum transfer 
$Q_{S}^2 = 0.1~{\rm (GeV/c)}^2$~\cite{sample01} 
\begin{equation}
G_{M}^{s} (Q_{S}^2)=+0.14 \pm 0.29~{\rm (stat)} 
\pm 0.31~{\rm (sys)}~{\rm n.m.}.
\label{data}
\end{equation}
On the other hand, baryons were described by topological 
solitons~\cite{skyrme61,witten83npb,brown83prl,hong98,hong02pr} and the MIT 
bag model~\cite{chodos74} was later unified with the Skyrmion model to yield the chiral 
bag model (CBM)~\cite{gerry791}, which then includes the pion cloud degrees of freedom and the 
chiral invariance consistently.  Moreover, the soliton  was exploited to yield 
superqualiton~\cite{hong01} in color flavor locking phase~\cite{rajagopal00}.

The QCD is the basic underlying theory of strong interaction, from which low energy hadron 
physics should be attainable.  Moreover, for hadron structure calculations, the 
coupling constant $g$ is not a relevant expansion parameter of QCD.  Long ago, 't Hooft noted 
that $1/N_{c}$ could be regarded as expansion parameter of QCD~\cite{thooft74} where $N_{c}$ 
is the number of colors and $gN_{c}^{2}$ is kept constant.  The properties of large $N_{c}$ limit 
of the QCD can be satisfied by the meson sector of the nonlinear sigma model such as the 
Skyrmion model.  

In this paper, we will study the chiral models such as the Skyrmion and chiral bag to 
yield theoretical predictions of proton strange form factor comparable to the recent 
experimental data of the SAMPLE Collaboration.  To do this, we will formulate the model 
independent sum rules for the proton strange form factor in terms of the baryon octet 
magnetic moments.  We will also investigate the Becci-Rouet-Stora-Tyutin (BRST) symmetries 
associated with the St\"uckelberg coordinates, ghosts and anti-ghosts involved in 
the Skyrmion model.

\section{BRST symmetries of Skyrmion in improved Dirac quantization }

Now, in order to study the hadron physics phenomenology, we treat $1/N_{c}$ as 
expansion parameter of QCD, so that the properties of large $N_{c}$ limit 
of the QCD can be satisfied by the SU(3) Skyrmion model whose 
Lagrangian is of the form~\cite{witten83npb}
\begin{equation}
L=\int{\rm d}^{3}x~\left[-\frac{f_{\pi}^{2}}{4}
{\rm tr}(l_{\mu}l^{\mu})+\frac{1}{32e^{2}}{\rm tr}[l_{\mu},l_{\nu}]^{2}\right]+L_{WZW}
\end{equation}
where $l_{\mu}=U^{\dagger}\partial_{\mu}U$ and $U\in$ SU(3) is described by 
pseudoscalar meson fields $\pi_{a}$ $(a=1,2,...,8)$ and the topological 
aspects can be included via the WZW action~\cite{witten83npb}.  Assuming maximal 
symmetry, we introduce the hedgehog ansatz $U_{0}$ embedded in the SU(2) 
isospin subgroup of SU(3) to yield the topological charge 
\begin{equation}
Q=-\frac{1}{2\pi}\chi_{E}(\theta-\sin\theta\cos\theta)=1
\label{charge}
\end{equation}
where $\theta$ is the chiral angle and $\chi_{E}$ is the Euler characteristic being 
an inter two in the spherical bag surface.  

In order to define the spin and isospin we can quantize, in the SU(2) Skyrmion for instance, 
the zero modes via 
\begin{equation}
U_{0}\rightarrow AU_{0}A^{\dagger}
\end{equation}
and 
\begin{equation}
A(t)=a^{0}+i\vec{a}\cdot\vec{\tau},
\end{equation}
with $a^{\mu}$ being the collective coordinates.  We can then 
obtain the Lagrangian 
\begin{equation}
L=-m_{0}+2i_{1}\dot{a}^{\mu}\dot{a}^{\mu}
\end{equation} 
where the static mass $m_{0}$ and the moment of inertia $i_{1}$ are calculable 
in the Skyrmion model.  Introducing the canonical momenta $\pi^{\mu}$ we can obtain 
the canonical Hamiltonian 
\begin{equation}
H=m_{0}+\frac{1}{8i_{1}}\pi^{\mu}\pi^{\mu}.
\end{equation}  
Note that the second-class geometrical constraints 
\begin{eqnarray}
\Omega_{1}&=&a^{\mu}a^{\mu}-1\approx 0,\nonumber\\
\Omega_{2}&=&a^{\mu}\pi^{\mu}\approx 0
\end{eqnarray}
should be treated via the Dirac 
brackets~\cite{dirac64}.  However, in the Dirac quantization, we have difficulties in 
finding the canonically conjugate pair, which were later overcome~\cite{bft} by 
introducing pair of auxiliary St\"uckelberg fields $\theta$ and $\pi_{\theta}$ with 
\begin{equation}
\{\theta, \pi_{\theta}\}=1.
\end{equation}  
In the Skyrmion the first-class constraints 
\begin{eqnarray}
\tilde{\Omega}_{1}&=&a^{\mu}a^{\mu}-1+2\theta,\nonumber\\
\tilde{\Omega}_{2}&=&a^{\mu}\pi^{\mu}-a^{\mu}a^{\mu}\pi_{\theta}
\end{eqnarray}
were constructed~\cite{hong99prd} to satisfy the strongly involutive 
Lie algebra 
\begin{equation}
\{\tilde{\Omega}_{1},\tilde{\Omega}_{2}\}=0.  
\end{equation}
Similarly, the first-class Hamiltonian 
was formulated to yield the baryon mass spectrum 
\begin{equation}
m_{B}=m_{0}+\frac{1}{2i_{1}}\left[J(J+1)+\frac{1}{4}\right]
\end{equation}
with the isospin quantum number $J$.
Here note that an additional global shift is due to the Weyl ordering correction.  Following the 
BRST quantization scheme~\cite{brst} with (anti)ghost and their Lagrangian multiplier fields, we 
obtain the BRST symmetric Lagrangian~\cite{hong99prd},
\begin{eqnarray}
L_{eff}&=&-m_{0}+\frac{2i_{1}\dot{a}^{\mu}\dot{a}^{\mu}}{1-2\theta} 
-\frac{2i_{1}\dot{\theta}^{2} }{(1-2\theta)^{2}}-\frac{\dot{\theta}\dot{b}}
{1-2\theta}\nonumber\\
& &-2i_{1}(1-2\theta)^{2}(b+2\bar{c}c)^{2}+\dot{\bar{c}}\dot{c}
\end{eqnarray}
invariant under the transformations, 
\begin{eqnarray}
\delta_{\lambda}a^{\mu}&=&\lambda a^{\mu}c,~~~
\delta_{\lambda}\theta=-\lambda(1-2\theta)c,\nonumber\\
\delta_{\lambda}\bar{c}&=&-\lambda b,~~~
\delta_{\lambda}c=\delta_{\lambda}b=0.
\end{eqnarray}
(For more details of the BRST quantization of the SU(2) and SU(3) Skyrmions, see Ref.~\cite{hong99prd} and 
Ref.~\cite{hong01prd}, respectively.)

\section{Model independent sum rules and proton strange form factors}

\begin{table}[t]
\caption{The baryon octet strange form factors}
\begin{center}
\begin{tabular}{lrrrr}
\hline
Input &$F_{2N}^{s}(0)$ &$F_{2\Lambda}^{s}(0)$ &$F_{2\Xi}^{s}(0)$ 
      &$F_{2\Sigma}^{s}(0)$\\ 
\hline
CBM  &0.30   &0.49 &0.25 &$-1.54$\\ 
Exp  &0.32   &1.42 &1.10 &$-1.10$\\ 
\hline\\
\end{tabular}
\end{center}
\end{table}

Next, we consider the CBM which is a hybrid of two different models: 
the MIT bag model at infinite bag radius on one hand and Skyrmion model at vanishing 
radius on the other hand.  (The explicit CBM Lagrangian is given in
Ref.~\cite{hong02pr} for instance.)  In the CBM the total topological
charge $Q$ in (\ref{charge}) is now splitted into the meson and
quark pieces to satisfy the Cheshire cat principle~\cite{ccp}.
Moreover, the quark fractional charge is given by sum of integer one 
(from valence quarks) and the quark vacuum contribution, which is also 
rewritten in terms of the eta invariant~\cite{atiyah}.  

In the collective quantization of the CBM, we explicitly obtain the
proton magnetic moment~\cite{hong93,hong93npa}
\begin{equation}
\mu_{p}=\frac{1}{90}(9I_{1}+24I_{2}+12I_{3}+16I_{4}-4I_{5}) 
+\frac{2I_{6}}{1125}\left(9I_{1}+4I_{2}-8I_{3}\right) 
\end{equation}
with the inertia parameters $I_{n}$ $(n=1,...,6)$ calculable in the CBM.  Similarly we 
construct the baryon octet magnetic moments to reproduce 
the Coleman-Glashow sum rules~\cite{coleman,hong93npa} such as $U$-spin symmetries, 
\begin{equation}
\mu_{\Sigma^{+}}=\mu_{p},~~~
\mu_{\Xi^{0}}=\mu_{n},~~~
\mu_{\Xi^-}=\mu_{\Sigma^{-}}.
\end{equation}

Now we define the Dirac and Pauli EM form factors via 
\begin{equation}
\langle p+q|\hat{V}^{\mu} |p\rangle = \bar{u}(p+q)\left[F_{1B}(q^2)
\gamma^{\mu}+\frac{i}{2m_B}F_{2B}(q^2)\sigma^{\mu\nu} 
q_\nu\right]u(p)
\end{equation}
where $q$ is momentum transfer and 
$\sigma^{\mu\nu}=\frac{i}{2}(\gamma^{\mu}\gamma^{\nu}-\gamma^{\nu}\gamma^{\mu})$ and 
$m_{B}$ is baryon mass.  The Sachs form factors are then given by 
\begin{equation}
G_{M}=F_{1B}+F_{2B},~~~
G_{E}=F_{1B}+\frac{q^{2}}{4m_{B}^{2}}F_{2B},
\end{equation}
so that, at zero momentum transfer, 
the Pauli strange form factor is identical to the Sachs strange form factor: 
\begin{equation}
F_{2B}^{s}(0) =G_{M}^{s}(0).
\end{equation}
In the SAMPLE experiment, they measured the neutral weak form factor 
\begin{equation}
G_{M}^{Z,p} = \left(\frac{1}{4}-\sin^{2} \theta_{W}\right)G_{M}^{p}-\frac{1}{4}G_{M}^{n}
-\frac{1}{4}G_{M}^{s}
\label{gemzp}
\end{equation}
with $G_{M}^p$ and $G_{M}^n$ being the proton and neutron Sachs form factors, to predict 
the proton strange form factor (\ref{data}) which is positive value contrary to the 
negative values from most of the model calculations except the predictions~\cite{hong93,hong97plb} 
of the SU(3) CBM and the recent predictions of the chiral quark soliton model~\cite{kim98} 
and the chiral perturbation theory~\cite{meissner00,vankolck}.  (See Ref.~\cite{hong02pr} for more details.) 

In the CBM the proton strange form factor is given by~\cite{hong93}
\begin{eqnarray}
F_{2N}^{s}(0)&=&\frac{1}{60}(21I_{1}-4I_{2}-2I_{3}-4I_{4}-2I_{5})\nonumber\\
& &+\frac{I_{6}}{2250}(-129I_{1}+76I_{2}-52I_{3})
\label{f2n0}
\end{eqnarray}
which, after some algebra with the other baryon octet strange form factors, yields 
the sum rule for the proton strange form factor in terms of the baryon octet magnetic 
moments only (for the other baryon sum rules see Ref.~\cite{hong01hep}) 
\begin{equation}
F_{2N}^{s}(0)=\mu_{p}-\mu_{\Xi^{-}}-(\mu_{p}+\mu_{n})
-\frac{1}{3}(\mu_{\Sigma^{+}}-\mu_{\Xi^{0}})
+\frac{4}{3}(\mu_{n}-\mu_{\Sigma^{-}}).
\label{f2n1}
\end{equation}
Explicitly calculating the inertia parameters $I_{n}$ numerically in (\ref{f2n0}), 
we predict the proton strange form factor, 0.30 n.m. as shown in Table 1.  Moreover, 
exploiting the experimental data for the baryon octet magnetic moments in (\ref{f2n1}) we obtain 
\begin{equation}
F_{2N}^{s}(0)=G_{M}^{s}(0)=0.32~{\rm n.m.}.  
\label{psff2}
\end{equation}
On the other hand, the quantities $G_{E,M}^Z$ in (\ref{gemzp}) for the 
proton can be determined via elastic parity-violating electron scattering to 
yield the experimental data $G_{M}^{s} (Q_{S}^2) = +0.14 \pm 0.29~{\rm (stat)} 
\pm 0.31~{\rm (sys)}$ n.m.~\cite{sample01} for the proton strange magnetic 
form factor.  Here one notes that the prediction for the proton strange form 
factor (\ref{psff2}) obtained from the sum rule (\ref{f2n1}) is comparable 
to the SAMPLE data and is shown in Table 1, together with those of the other 
baryon strange form factors.  Moreover, from the relation (\ref{gemzp}) at zero 
momentum transfer, the neutral weak magnetic moment of the nucleon can be 
written in terms of the nucleon magnetic moments and the proton strange form 
factor~\cite{mckeown002}
\begin{equation}
4\mu_{p}^{Z}=\mu_{p}-\mu_{n}-4\sin^{2}\theta_{W}\mu_{p}-F_{2N}^{s}(0).
\label{mupz}
\end{equation}

\section{Conclusions}

In conclusion, we discussed the SAMPLE experiments in the topological solitons such as 
the Skymion and chiral models to predict baryon strange form factors by constructing the 
model independent sum rules for the proton strange form factor in terms of the baryon 
octet magnetic moments.  We also exploited the improved Dirac quantization scheme to 
investigate the BRST symmetries associated with the St\"uckelberg coordinates, ghosts 
and anti-ghosts involved in the Skyrmion model.

\end{document}